# Effect of Substrate Modes on Thermal Transport in Supported Graphene

Zhun-Yong Ong[1,2] and Eric Pop[1,3,4]

[1]*Micro and Nanotechnology Lab, University of Illinois, Urbana-Champaign, IL 61801, USA*

[2]*Dept. of Physics, University of Illinois, Urbana-Champaign, IL 61801, USA*

[3]*Dept. of Electrical & Computer Eng., University of Illinois, Urbana-Champaign, IL 61801, USA*

[4]*Beckman Institute, University of Illinois, Urbana-Champaign, IL 61801, USA*

We examine thermal transport in graphene supported on $SiO_2$ using molecular dynamics (MD) simulations. Coupling to the substrate reduces the thermal conductivity (TC) of supported graphene by an order of magnitude, due to damping of the flexural acoustic (ZA) phonons. However, increasing the strength of the graphene-substrate interaction *enhances* the TC of supported graphene, contrary to expectations. The enhancement is due to the coupling of graphene ZA modes to the substrate Rayleigh waves, which linearizes the dispersion and increases the group velocity of the hybridized modes. These findings suggest that the TC of 2-D supported graphene is tunable through surface interactions, providing a novel possibility for controlled energy flow in nanomaterials.

*Contact: zong2@illinois.edu, epop@illinois.edu

## I. INTRODUCTION

As a result of its outstanding electrical[1, 2] and thermal properties,[3, 4] single layer graphene (SLG) has become a promising nanomaterial for future electronics. In this context, graphene will be supported by and integrated with insulators such as $SiO_2$, both for circuit and heat spreader applications.[5] Thus, thermal energy flow will be limited both by the thermal conductivity (TC) of the supported graphene[6-8] and by the thermal boundary conductance (TBC) at the graphene-$SiO_2$ interface.[9-11] Importantly, interaction with the substrate can modify the thermo-mechanical properties of the graphene. For instance, to explain the reduced TC of $SiO_2$-supported graphene, Refs. 6 and 7 suggested that the flexural acoustic (ZA) phonons are responsible for ~70% of the TC in free graphene, and are damped when graphene is placed on a substrate. On the other hand, Refs. 3 and 12 concluded (albeit contrastingly[13]) that ZA phonons do not make a substantial contribution to the TC of free graphene, on account of their low group velocity and large Grüneisen parameter. Moreover, studies that did treat the substrate interaction[6, 7] had considered it as a mechanism akin to interface and roughness scattering, or as an on-site static perturbation,[14, 15] but did not take into account the complete degrees of freedom of the substrate material.

In this work, we use both molecular dynamics (MD) simulations and continuum modeling to elucidate the heat flow mechanisms in supported and suspended graphene. We find that coupling with an $SiO_2$ substrate reduces the TC of supported graphene by an order of magnitude compared to suspended graphene due to damping of the ZA modes, consistent with the experiments.[6] However, the TC of supported graphene can be tuned through the graphene-$SiO_2$ interaction and, unexpectedly, we find that the TC is enhanced when this coupling increases. This result is counterintuitive at first sight, but may be explained by the coupling of ZA modes in the graphene with Rayleigh waves of the $SiO_2$. These results provide crucial and novel directions for the tuning of nanostructure thermal properties through carefully controlled environmental interactions.

## II. SIMULATION METHODOLOGY

We use the LAMMPS MD package for our thermal transport simulations.[16] As in our earlier work,[17, 18] we employ the REBO potential,[19, 20] to model the atomic interactions between the C atoms and the Munetoh potential[21] to model the atomic interactions between the Si-O atoms. The C-Si and C-O couplings are modeled as van der Waals (vdW) interactions using the Lennard-Jones (LJ) potential $V_{ij}(r) = 4\chi\varepsilon_{ij}[(\sigma_{ij}/r)^{12}-(\sigma_{ij}/r)^{6}]$ where $i$ = C, $j$ = Si or O, and $r$ is the inte-ratomic distance. The standard numerical values of the LJ parameters $\varepsilon_{ij}$ and $\sigma_{ij}$ are given in Ref. 17. The dimensionless parameter $\chi$ represents the relative strength of the van der Waals coupling, with a default value $\chi = 1$.

Our simulation domain includes a maximum of 59904 atoms, and the largest dimension is $L_z = 295.2$ Å in the direction of heat flow, as shown in Fig. 1. As the size of the system is limited, the temperature gradients can be relatively large (~1 K/nm). However, comparable thermal gradients have been used in MD simulations of interfacial thermal resistance[22, 23] or other thermal transport phenomena,[24-26] with meaningful results. Even higher thermal gradients (>10 K/nm) have been sustained in experiments investigating the thermal coupling of heated atomic force microscope (AFM) tips with films of polystyrene[27] and individual carbon nanotubes.[28]

## III. SIMULATION RESULTS

Recent work has established that isolated (freely suspended) monolayer graphene[3, 4, 29-31] has a high TC of several thousand $Wm^{-1}K^{-1}$. However, it has also been suggested that a substrate in contact with the SLG can affect (dampen) the propagation of SLG phonons and reduce its TC.[6-8] To study the effect of this interaction, we compare thermal transport along (i) an isolated graphene sheet in Fig. 1(a), (ii) an $SiO_2$ thin film as in Fig. 1(b), and (iii) a graphene-$SiO_2$-graphene sandwich as in Fig. 1(c). In (i) and (iii), the graphene is oriented such that the 'zigzag' pattern runs parallel to the direction of heat flow.

### A. Free graphene

We compute the TC of free graphene using equilibrium MD (EMD) and non-equilibrium MD (NEMD) simulations. In the NEMD simulation, we evolve the system using a time step of 0.2 fs while thermostatting the atoms in the center at 310 K and the edge atoms at 290 K. The thermostatting of the atoms is carried out by rescaling the selected atoms at regular time intervals of $\Delta t = 4$ fs. At each time interval, the velocity rescaling leads to a kinetic energy change $\varepsilon$ of

$$\varepsilon = \sum_{\alpha} \frac{1}{2} m_{\alpha} \left( \mathrm{v}_{\alpha,\mathrm{f}}^2 - \mathrm{v}_{\alpha,\mathrm{i}}^2 \right) \tag{1}$$

where $m_\alpha$ is the mass of the $\alpha$-th atom and $\mathrm{v}_{\alpha,\mathrm{i}}$ and $\mathrm{v}_{\alpha,\mathrm{f}}$ are its initial and final velocities. The constant application of thermostatting forces the system out of equilibrium and thus creates an aver-age net thermal current from the center to the edge. We find that this constant thermostatting leads to a steady-state heat current and that the time-averaged $\varepsilon$ of the middle atoms is equal to the negative of the time-averaged ε of the edge atoms. At steady state, we compute the energy transfer rate $Q_G$ and the heat flux $Q_z = Q_G/(2A)$ where $A$ is the cross-sectional area (174 Å$^2$). The energy transfer rate $Q_G$ is obtained from the gradient of the linear plot of $E(t)$ versus $t$, where $E(t)$ is the cumulative sum of the kinetic energy change $\varepsilon$ over a time interval $t = 1$ ns. The temperature profile is also averaged over 1 ns. At steady state, the average heat flux is $Q_z = 1.25 \times 10^{11}$ Wm$^{-2}$ and the energy transfer rate between the middle and the edge atoms is $Q_G = 4.37 \times 10^{-7}$ W.

The temperature profile from the center to the edge of the 'free' graphene is shown in Fig. 2(a). This profile is nonlinear, with temperature 'slips' near the thermostats due to finite size effects[32] and the high TC of graphene. Such temperature profiles have also been noted in carbon nanotube (CNT) simulations.[26] Thus, some care must be taken in extracting the TC from this nonlinear profile which indicates that thermal transport is not fully diffusive. A naïve computation of the TC using the formula $\kappa_G = -Q_z(\Delta T/\Delta z)^{-1}$ would give a value of 92 Wm$^{-1}$K$^{-1}$. To obtain the correct diffusive TC we use the temperature gradient of the middle portion between the thermostats[26] to avoid edge effects, obtaining $\kappa_G = -Q_z(dT/dz)^{-1} = 256$ Wm$^{-1}$K$^{-1}$ for 'free' graphene.

To complement the NEMD calculation of the TC, we also compute the TC from equilibrium MD simulations using the Green-Kubo relation[33]

$$\kappa_{zz} = \lim_{t\to\infty} \frac{1}{L_x L_y L_z k_B T^2} \int_0^t dt' \langle Q_z(t) Q_z(0) \rangle \tag{2}$$

where $L_x$, $L_y$, $L_z$, $k_B$ and $T$ are the width, height, length, the Boltzmann constant and the temperature (300 K) respectively. The instantaneous value of the heat flux was recorded over 5 ns at intervals of 5 fs to produce a time series data which we use to compute $\kappa_{zz}$ as in Eq. (2). The integral of the heat flux autocorrelation is shown in Fig. 2(b) and found to converge to $\kappa_{zz} \approx 248$ Wm$^{-1}$K$^{-1}$, in excellent agreement with our estimate using NEMD. This also validates the graphene TC obtained from the temperature gradient away from the edges in the NEMD simulation.

Our free graphene TC value is smaller than that by Evans et al.[33] but is in close agreement with those reported by Guo et al.[34] We attribute the discrepancy to the choice of interatomic potential for the C atoms. Evans et al. used the Tersoff potential[35] while we used the RE-BO/Brenner potential[19, 20] as Guo et al. did. Lower TC values in the range of several hundred $Wm^{-1}K^{-1}$ have also been obtained for MD simulations of CNTs[36-38] using the REBO/Brenner potential. This discrepancy is not significant for our current study because our main intention is not to present another MD calculation of the graphene TC, but to study the effect of the substrate coupling on thermal transport in graphene.

We also comment on the role that size effects may play given the finite size of our simulated system (Fig. 1). A comparison with recent measurements of TC for isolated monolayer graphene[4, 29-31] (~1800-5150 $Wm^{-1}K^{-1}$) and with the highest reported values for pyrolytic graphite[39] (~2000 $Wm^{-1}K^{-1}$) shows that our simulated TC for free graphene tends to be smaller. We surmise that the difference between the measured and the simulated TC values is in part due to the limitation of our simulation domain size[40, 41] and also in part to the choice of interatomic potential, as mentioned earlier. However, while some boundary effects are seen in Fig. 2(a), it is apparent that they do not change the influence of the substrate on the graphene. Like Refs. 24, 38 and 42, one is able to find *changes* to thermal conduction caused by environmental or structural modifications even when the simulation domains are relatively small. Ideally, the effect of length on the thermal conduction behavior of supported graphene should be studied further, but the limitations of our computational resources prevent us from doing so at the time.

### B. Isolated $SiO_2$ thin film

For the $SiO_2$ film shown in Fig. 1(b), as with free graphene, we use both EMD and NEMD methods to calculate the TC. The conditions for the NEMD simulation of $SiO_2$ are identical to those of free graphene. However, unlike in free graphene, we obtain a linear temperature profile [see Fig. 2(c)] at steady state due to the very low TC of $SiO_2$, showing that heat conduction is in the diffusive regime. The average heat flux and TC obtained are $Q_z = 1.64 \times 10^9$ $Wm^{-2}$ and $\kappa_{ox} = 1.17$ $Wm^{-1}K^{-1}$ respectively from the NEMD simulation. The average energy transfer rate between the middle and edge atoms is $Q_{ox} = 8.11 \times 10^{-8}$ W, defined as the rate of energy transfer between the thermostatted Si/O atoms like $Q_z$ is. The TC of the $SiO_2$ block is also computed using Eq. (2), with the integral of the heat flux autocorrelation shown in Fig. 2(d), giving an average $\kappa_{ox} \approx 1.21$ $Wm^{-1}K^{-1}$, once again in good agreement with and validating the result from the NEMD simulation.

### C. Sandwich Structure

We use the sandwich structure shown in Fig. 1(c) to simulate thermal transport along supported graphene with NEMD. At steady state, we obtain a linear temperature profile similar to that in Fig. 2(c), and the average energy transfer rate from middle to edge atoms is $Q = 1.62 \times 10^{-7}$ W or about double $Q_{ox}$ but smaller than $Q_G$. To determine the relative contribution of the graphene layers in the sandwich, we subtract $Q_{ox}$ from $Q$ and obtain an effective energy transfer rate $Q_{sG} = 4.05 \times 10^{-8}$ W per supported SLG sheet, assuming the TC of the $SiO_2$ is not altered by its interaction with the graphene. This shows a 90.7% reduction in thermal transport of supported SLG (vs. the freely suspended case, $Q_G = 4.37 \times 10^{-7}$ W above) as a result of contact with the substrate. A similar phenomenon was also seen in MD simulations of a CNT on Si substrate.[43]

This reduction was recently measured experimentally in supported SLG[6] and was attributed to the suppression of the dominant phonon ZA modes by the substrate.[7] On the other hand, it has also been suggested that ZA modes do not contribute significantly to the TC.[3, 12] To understand the damping effect of the substrate on the acoustic modes in SLG in the context of our MD simulations, we turn to Fig. 3 where we compute the spectral energy density[44] of a 1920-atom 98.4 × 51.5 Å SLG sheet with and without the substrate (see the supplementary information for details of the computation). The SLG has 40 repeating units in the z direction which give us 20 distinct wave vector points (in the z-direction) in the spectrum, enumerated from $k = 1$ to 20 with wavelength inversely proportional to $k$. We show the spectra for $k = 1$ to 4 of the LA (longitudinal acoustic) and ZA modes in Figs. 3(a) and 3(b), respectively. The LA peaks in Fig. 3(a) remain sharp and distinct with and without the substrate, indicating weak substrate coupling and ruling out the possibility that the >90% reduction in TC is due to damping of the in-plane acoustic modes. On the other hand, the significant broadening and frequency shift of the ZA peaks in supported graphene indicate strong coupling to the substrate [Fig. 3(b)]. The broadening in free graphene is smaller at low wave vector numbers, suggesting that the long wavelength ZA modes have the longest lifetimes and dominate thermal transport. This finding confirms the role of the substrate in damping the ZA modes and the TC of supported graphene.

## IV. DEPENDENCE OF THERMAL CONDUCTION ON SUBSTRATE INTERACTION

Given that the damping of the ZA modes reduces the TC of supported graphene, then it would seem that increasing the strength of the interfacial interaction should lead to further reduc-tion of TC. To verify this hypothesis, we redo our simulation of thermal transport along the sandwich structure for different values of χ, where χ scales the strength of the vdW interaction between graphene and $SiO_2$, and χ = 1 corresponds to the original parameter values.[17] We plot the flux going through the two supported graphene sheets as $Q - Q_{ox}$ versus χ in Figs. 3(c) and 3(d) for χ = 0.1 – 10. For comparison, we also plot $2Q_G$ for the case of χ = 0 as the heat flux through two freely suspended graphene sheets. Surprisingly, for χ > 0 we find that as χ increases, the heat flux carried by the supported graphene sheets ($Q - Q_{ox}$) also *increases*, even though we expected thermal conduction through the SLG to decrease as a result of stronger damping of the ZA phonons. In fact, we find that the heat flux carried by the $SiO_2$-supported SLG increases by a factor of three as χ varies from 0.1 to 10.

We comment here on the possibility of tuning the graphene-substrate interaction in practice. For instance, it has been found that the binding energy between C atoms and the substrate surface varies widely, depending on the atomic species present on the surface [45]. Ab initio calculations suggest[45] that the binding energy from an O-terminated surface is more than an order of magnitude larger than that from a Si-terminated surface. Thus, it is possible or even probable that the strength of the coupling between the graphene and the substrate can vary widely, depending on the specific atomic composition of the surface. It is known that the O/Si ratio on the surface of an $SiO_2$ film can be altered via physical sputtering.[46] This suggests that the graphene-$SiO_2$ interaction itself could be 'tuned' in practice.

## V. POSSIBLE MODE HYBRIDIZATION BETWEEN GRAPHENE AND SUBSTRATE

To obtain a better physical understanding of the counterintuitive result obtained numerically above, we examine how contact with the substrate modifies the flexural acoustic waves, the continuum limit of the long wavelength ZA modes which dominate thermal transport in the SLG. We first examine the dispersion relation in a continuum model[47] of an ideal graphene membrane supported on an isotropic elastic substrate. The notation in the following discussion follows that of Refs. 48 and 49. The equation of motion of the membrane is

$$u_0(\boldsymbol{q},\omega) = -M_0(\boldsymbol{q},\omega)\,\sigma(\boldsymbol{q},\omega) \tag{3}$$

where $u_0$ is the out-of-plane displacement, $\boldsymbol{q}$ is the wave vector parallel to the interface, ω is the radial frequency and $M_0(\boldsymbol{q},\omega) = 1/(\kappa q^4-\rho_0\omega^2-i0)$ is the response function to the surface stress $\sigma(\mathbf{q},\omega)$. $\rho_0$ and $\kappa$ are respectively the areal mass density and the bending stiffness of the membrane. The equation of motion for the substrate surface can be written as

$$u_1(\boldsymbol{q},\omega) = M_1(\boldsymbol{q},\omega)\sigma(\boldsymbol{q},\omega) \tag{4}$$

where, following Appendix A in Ref. 48, $M_1(q,\omega) = \left(i/\rho_1 c_T^2\right) p_L(q,\omega)/S(q,\omega)\left(\omega/c_T\right)^2$, $S(q,\omega) = \left[\left(\omega/c_T\right)^2 - 2q^2\right]^2 + 4q^2 p_T p_L$, $p_L(q,\omega) = \left[\left(\omega/c_L\right)^2 - q^2 + i0\right]^{1/2}$ and $p_T(q,\omega) = \left[\left(\omega/c_T\right)^2 - q^2 + i0\right]^{1/2}$. The $i0$ term in $p_L$ and $p_T$ implies the square root function has its branch cut along the negative real axis; $c_T$ = 3743 m/s, $c_L$ = 5953 m/s and $\rho_1$ = 2200 kg/m$^3$ are respectively the transverse and longitudinal sound velocities and the mass density in the substrate with their values taken from Ref. 47. The interfacial stress is defined as $\sigma(\boldsymbol{q},\omega) = K[u_0(\boldsymbol{q},\omega) - u_1(\boldsymbol{q},\omega)]$ where $K = 1.82 \times 10^{20}$ N/m$^3$ is the interfacial spring constant. Combining this with Eqs. (3) and (4), the dispersion relation of the composite membrane-substrate system is determined by the zeros of the equation

$$h(\boldsymbol{q},\omega) = 1 + K\left[M_0(\boldsymbol{q},\omega) + M_1(\boldsymbol{q},\omega)\right]. \tag{5}$$

We plot $(d/d\omega)\mathrm{Re}[1/h(\boldsymbol{q},\omega)]$ in Fig. 4 for different values of the coupling strength χ. At the zeros of $h(\mathbf{q},\omega)$, $(d/d\omega)\mathrm{Re}[1/h(\boldsymbol{q},\omega)]$ diverges and we see the corresponding fine black lines. In Fig. 4(a), we take χ = 0.0001 (weak coupling, nearly free SLG) and we observe one linear and one quadratic curve, corresponding to the substrate Rayleigh modes (one with group velocity ~0.9 $c_T$) and the graphene flexural acoustic waves, respectively. At small wave vector $q$, the group velocities of the flexural modes are small due to their quadratic dispersion. However, as we increase the graphene-$SiO_2$ interaction $\chi$, [Figs. 4(b) to (d)] we observe the formation of two hybridized non-dispersive modes with higher group velocities, one corresponding to ~$c_L$ and the other ~0.8 $c_T$. When $K$ is large, the group velocity of the Rayleigh wave goes from 0.9 $c_T$ to $c_L$ for large $q$ values because the zeros of Eq. (5) are approximately given by the zeros of $M_1$. We note that this continuum model is not necessarily fully accurate for the flexural motion of the graphene. Nevertheless, it illustrates the change in the dispersion relation arising from the coupling of a membrane to a solid substrate.

Thus, the enhanced thermal conduction through supported SLG with increased substrate coupling can be explained by the formation of higher-velocity elastic waves at the graphene-$SiO_2$ interface. Although ZA modes are initially damped when graphene is placed on a substrate, with increased interfacial coupling the flexural motion becomes coupled to Rayleigh waves of the substrate. When the coupling is very strong, the dispersion relation in fact becomes nearly *linear* [as can be seen in Fig. 4(d)] and the long wavelength modes propagate at higher group velocities, leading to the enhancement of thermal transport along the interface.

## VI. CONCLUSIONS

In summary, we used MD simulations to provide evidence that thermal transport in free graphene is dominated by ZA modes; in addition, we found an order of magnitude reduction in the TC of supported graphene due to damping of ZA modes by the substrate (here, $SiO_2$). However, when the graphene-substrate vdW coupling is increased, heat flow along supported graphene increases as well, and can be modulated by up to a factor of three. This is a counterintuitive result, yet it can be explained by noting that strong coupling between quadratic ZA modes of graphene with surface waves of the substrate leads to a hybridized linear dispersion and higher phonon group velocity. These results highlight an interesting and novel route for tuning thermal transport in 2D nanostructures like graphene, via carefully controlled environmental interactions.

After submission, it was brought to our attention that a similar simulation study of TC enhancement through substrate coupling was performed by Guo et al.[50] In their paper, using a simple MD example, they show that the heat flux through a pair of coupled chains can be enhanced beyond that through the isolated chains, provided that the phonon resonance is minimized. According to them, the modulation of this resonance can be achieved by choosing a substrate with a very different atomic mass and/or spring constant. Although we also propose that the TC can be modulated by substrate coupling, our paper takes a fundamentally different approach and argues that some thermal transport enhancement is achieved through the hybridization of the long wavelength flexural modes in the graphene with the faster Rayleigh waves of the substrate, by strengthening the interaction between the graphene and the substrate. Furthermore, the physics in our paper is that of a 2-dimensional structure supported on a semi-infinite 3-dimensional substrate whereas the thermal transport enhancement claimed by Guo et al.[50] is only shown for 1-dimensional or pseudo-1-dimensional structures (such as CNTs). Nevertheless, the similarity is intriguing and suggests that one may be able to control thermal transport in low-dimensional structures by modulating their coupling to substrates.

**ACKNOWLEDGEMENTS**

This work has been partly supported by the Nanoelectronics Research Initiative (NRI), the NSF under Grant No. CCF 08-29907 and a gift from Northrop Grumman Aerospace Systems (NGAS). We acknowledge valuable technical discussion with R. L. Weaver and J. Shiomi, computational support from R. Toghraee and U. Ravaioli, and computational resources from the Texas Advanced Computing Center (TACC) at UT Austin.

## APPENDIX: SPECTRAL ENERGY DENSITY

The method of spectral energy density (SED)[44] allows us to group phonon modes by translational symmetries and to construct a spectrum of discrete peaks corresponding to those phonon modes. To compute the SED, we simulated a 1920-atom 98.4 × 51.5 Å SLG sheet, with and without the substrate, at constant energy and volume for $\tau$ = 1.2 ns after an equilibration period of 0.6 ns at constant temperature (300 K), and recorded the velocity ($v_x$, $v_y$ and $v_z$) trajectory of each atom at intervals of 5 fs. To convert the atomic velocity trajectories into the spectral energy density, we use the translational symmetry in the x and z-directions by defining a 4 basis atom unit cell as shown in Fig. 5. The spectral energy density is then given by the expression

$$\Phi(k_z,k_x,\nu)=\left\langle \sum_{b=1}^{4}\sum_{\alpha}\frac{m}{2}\left|\frac{1}{N_zN_x\tau}\sum_{n_z=0}^{N_z-1}\sum_{n_x=0}^{N_x-1}\exp\left[2\pi i\left(\frac{k_zn_z}{N_z}+\frac{k_xn_x}{N_x}\right)\right]\int_0^{\tau}dt\,\exp(2\pi i\nu t)\mathrm{v}_{\mathrm{b},\alpha}(n_z,n_x,t)\right|^2\right\rangle \quad (6)$$

where $k_z$, $k_x$, $\nu$, $m$, $n_z$, $n_x$, $N_z$ and $N_x$ are the z-direction wave vector, the x-direction wave vector, the frequency, the C atom mass, the unit cell x-direction position (from 0 to 11) and the unit cell z-direction position (from 0 to 39); $\mathrm{v}_{\mathrm{b},\alpha}$ is the velocity of the $b$-th basis atom in the $\alpha$-th direction. For computational convenience, we consider only the wave vector points in the z-direction and set $k_x$ = 0. We also drop the subscript $z$ in $k_z$. Thus, Eq. (6) becomes

$$\Phi(k,\nu)=\left\langle \sum_{b=1}^{4}\sum_{\alpha}\frac{m}{2}\left|\frac{1}{N_zN_x\tau}\sum_{n_z=0}^{N_z-1}\sum_{n_x=0}^{N_x-1}\exp\left(\frac{2\pi ikn_z}{N_z}\right)\int_0^{\tau}dt\,\exp(2\pi i\nu t)\mathrm{v}_{\mathrm{b},\alpha}(n_z,n_x,t)\right|^2\right\rangle \quad (7)$$

which, under the Lorentzian approximation that the time autocorrelation of the normal mode coordinates decays exponentially, reduces to a sum of 12 Lorentzian functions

$$\Phi(k,\nu)=\sum_{\beta=1}^{12}\frac{I_\beta(k,\nu)}{\left[\nu-\nu_\beta(k)\right]^2+\frac{1}{4}\gamma_\beta(k)^2} \quad (8)$$

where $\beta$ indexes the phonon branch, and $I_\beta(k_z,\nu)$, $\nu_\beta(k_z)$ and $\gamma_\beta(k_z)$ are the intensity, eigenfrequency and lifetime of the corresponding phonon mode. Using the SED method, we are able to observe discrete Lorentzian peaks in the spectrum with their full-width half-maximums corresponding to their lifetimes (Fig. 3). When the modes are coupled to external degrees of freedom, they undergo additional broadening, and if the coupling is strong enough such that the Lorentzian approximation no longer holds, the Lorentzian shape is lost.

As the acoustic modes are primarily responsible for thermal transport, we plot the peaks corresponding to the LA and ZA modes in Fig. 3. To plot the LA modes in Fig. 3(a), we first assumed that they are purely longitudinally polarized (in the z-direction) and computed Eq. (7) using only the z-component of the atomic velocities, with $k = 1$ to 4, in free graphene. The 4 peaks (out of a total of 48) with the lowest frequencies were taken to be the LA modes. When we repeated the calculation with the x-component of the atomic velocities, the 4 peaks with the lowest frequencies (TA modes for $k = 1$ to 4) had positions different from those of the LA modes. The same was also true when we repeated the calculation with the y-components of the atomic velocities to get the ZA modes in Fig. 3(b). This indicates that, for $k = 1$ to 4, the LA, TA and ZA modes are purely polarized in the longitudinal, in-plane transverse and out-of-plane transverse directions respectively. In supported graphene, the corresponding LA and TA peaks can still be distinguished but the ZA peaks are shifted and broadened as shown in Fig. 3(b).

Figures 3(a) and 3(b), plot the LA and ZA peaks for $k = 1$ to 4. The peaks for the LA modes in free graphene in Fig. 3(a) are sharp Lorentzians; the peaks for the same modes in supported graphene have noticeable broadening but still retain their Lorentzian shape, which indicates that the coupling to the substrate is relatively weak. In contrast, the ZA peaks are Lorentzian in free graphene but, except the $k = 1$ peak, lose their Lorentzian profile in supported graphene, suggesting that the phonon modes are so strongly coupled to the substrate that the Lorentzian approximation does not hold. In this case, the coupling to the substrate can no longer be approximated as a weak perturbation and almost pure ZA modes cannot be isolated because they have hybridized with the substrate modes.

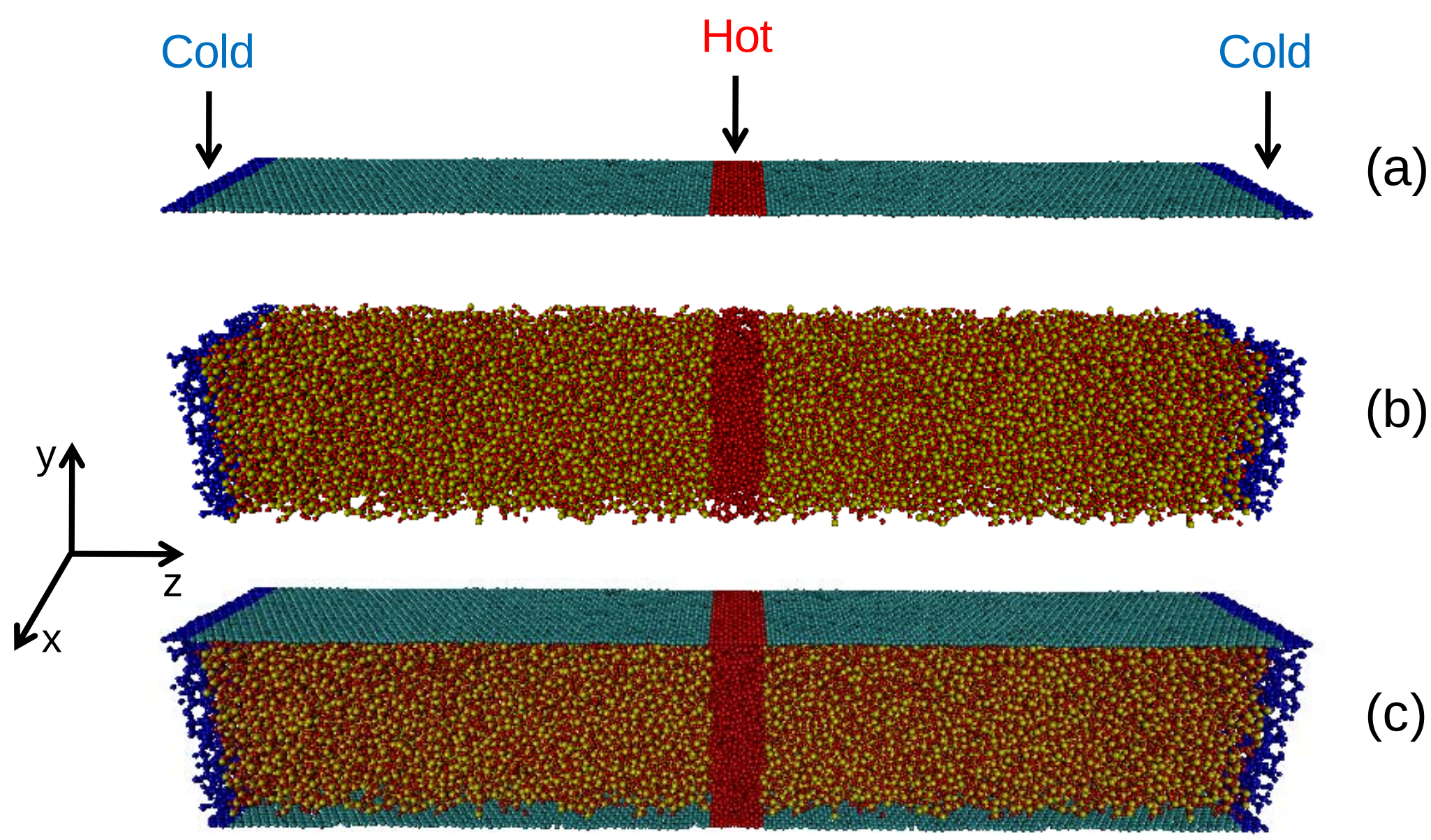

FIG. 1. (Color online) Set up for simulation of thermal transport in free and supported graphene in the z-direction. The graphene is coplanar to the x-z plane, and $L_x$, $L_y$ and $L_z$ are the width, height and length of the simulated structure, respectively. Periodic boundary conditions are imposed in the x and z directions. We simulate thermal transport in (a) SLG sheet ($L_y = 3.4$ Å), (b) $SiO_2$ thin film ($L_y = 48.0$ Å) and (c) sandwich structure with the $SiO_2$ film sandwiched between two SLG sheets ($L_y = 54.8$ Å). $L_x = 51.1$ Å and $L_z = 295.2$ Å for all structures. We impose periodic boundary conditions in the *x*- and *z*-direction. In the NEMD simulations, the 'hot' atoms at the center are thermostatted at 310 K, and the 'cold' edges are at 290 K.

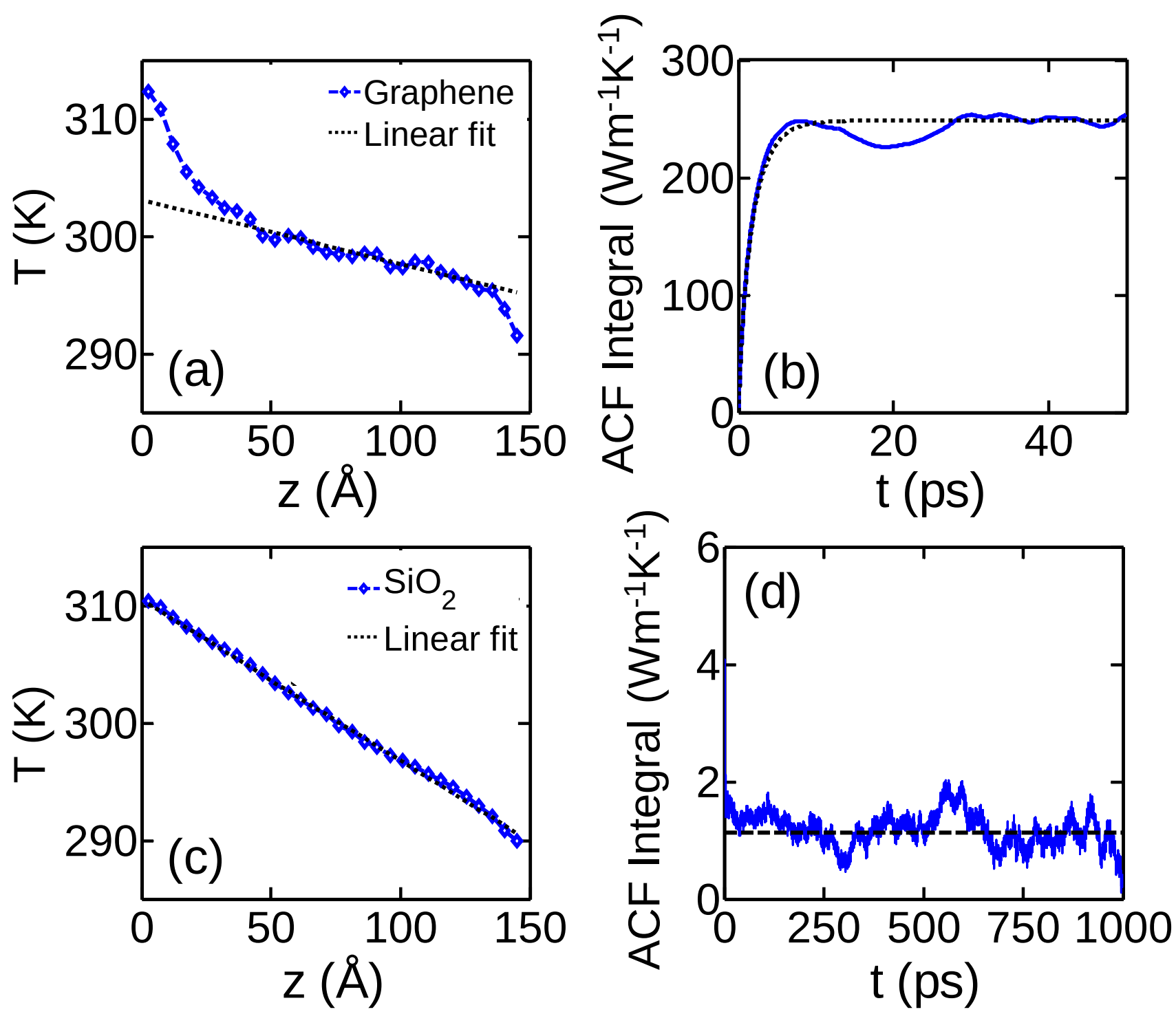

FIG. 2. (Color online) (a) Temperature profile from center to edge of free graphene, and (b) integral of the heat flux autocorrelation function (ACF). (c) Temperature profile from center to edge of $SiO_2$ film and (d) integral of the heat flux ACF [Eq. (2)].

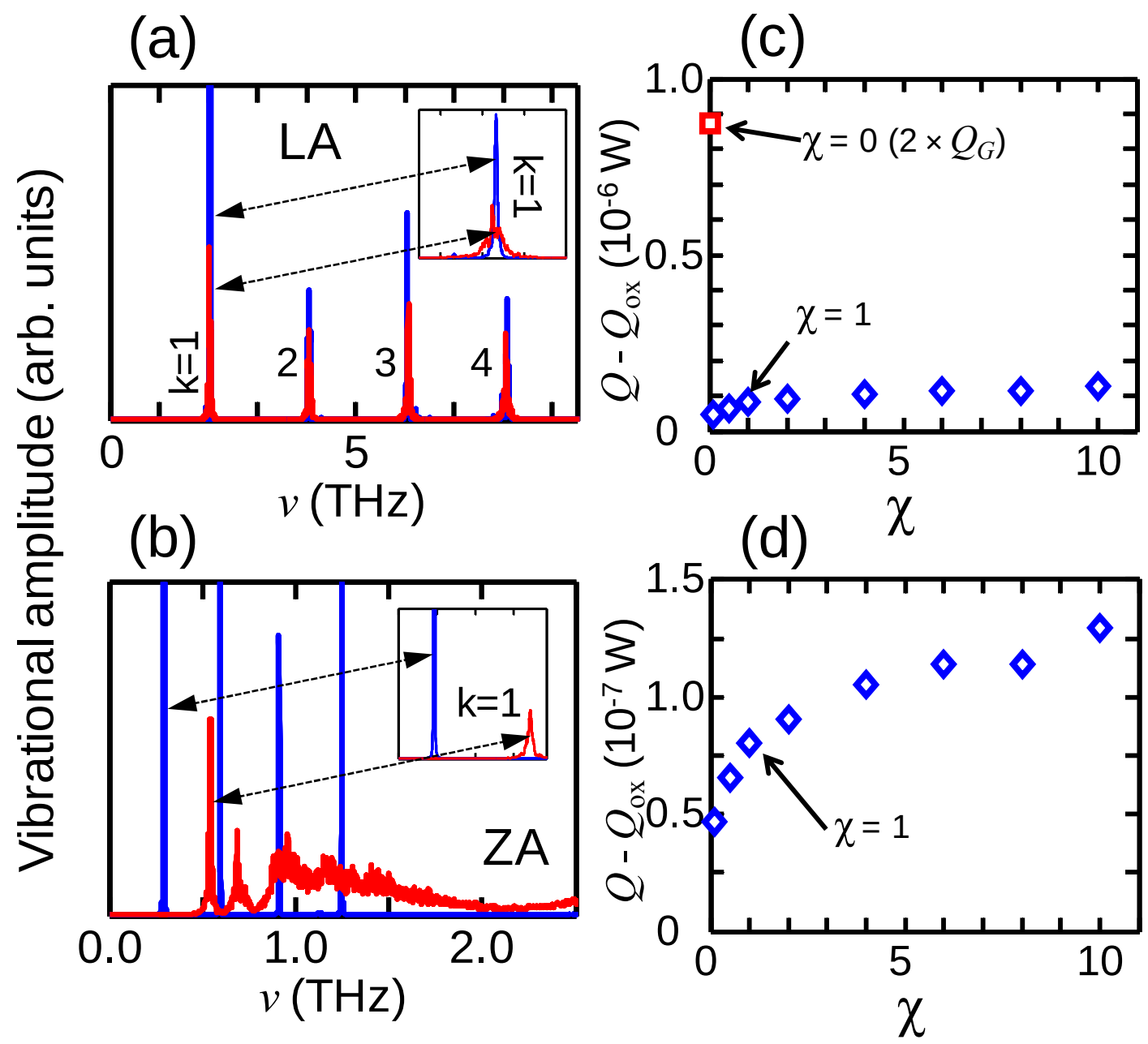

FIG. 3. (Color online) (a)-(b) Long wavelength (k = 1 to 4) vibrational spectrum of free graphene (taller, narrower peaks) and supported graphene (shorter, broader peaks). (a) Distinct LA phonon peaks show little effect of the substrate. (b) ZA phonon peaks are significantly broadened and upshifted after the graphene is supported on the substrate. (c) Plot of heat flux through two suspended ($2Q_G$) and supported ($Q - Q_{ox}$) graphene layers versus $\chi$, which scales the van der Waals interaction at the graphene-$SiO_2$ surface; $\chi = 1$ is the default interaction strength. The heat flux through supported graphene is approximately one order of magnitude lower than that through the suspended SLG. (d) Zoom-in version of previous plot. The heat flux along supported SLG ($Q - Q_{ox}$) increases by a factor of three as the graphene-$SiO_2$ interaction $\chi$ increases from 0.1 to 10. This counterintuitive result can be explained by the coupling of graphene ZA modes with $SiO_2$ surface waves, which increase the group velocity of the hybridized modes (see text and Fig. 4).

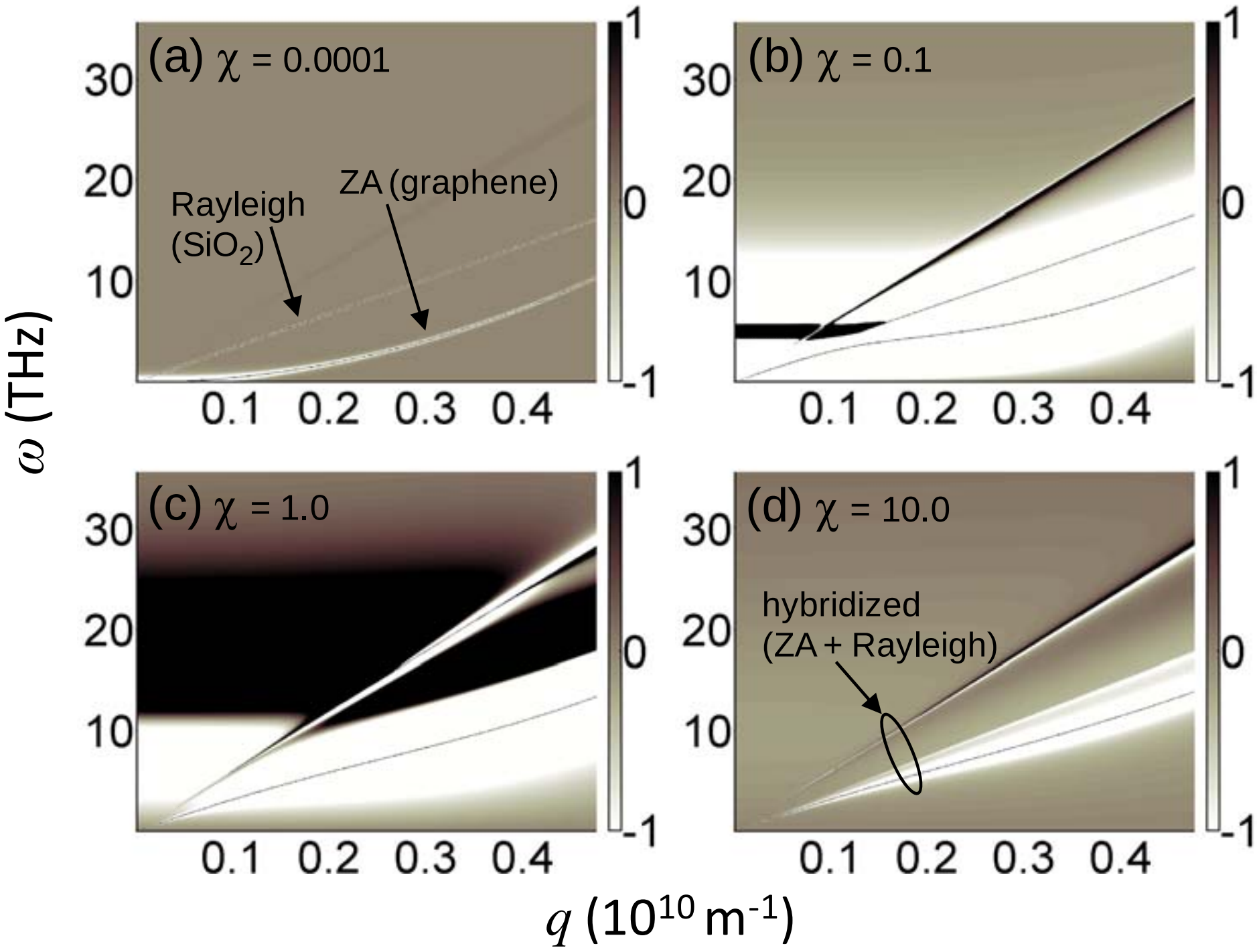

FIG. 4. (Color online) Calculated $(d/d\omega)\,\mathrm{Re}[1/h(\boldsymbol{q},\omega)]$ for different values of the graphene-substrate interaction $\chi$, from (a) nearly-free graphene ($\chi = 0.0001$), (b) $\chi = 0.1$, (c) $\chi = 1$ (default case), and (d) ten-fold stronger interaction, $\chi = 10$. For nearly free graphene the ZA modes show the typical quadratic dispersion. As the graphene-substrate interaction increases, ZA graphene modes hybridize with $SiO_2$ Rayleigh waves, leading to linearized dispersion, higher group velocity, and enhanced thermal transport.

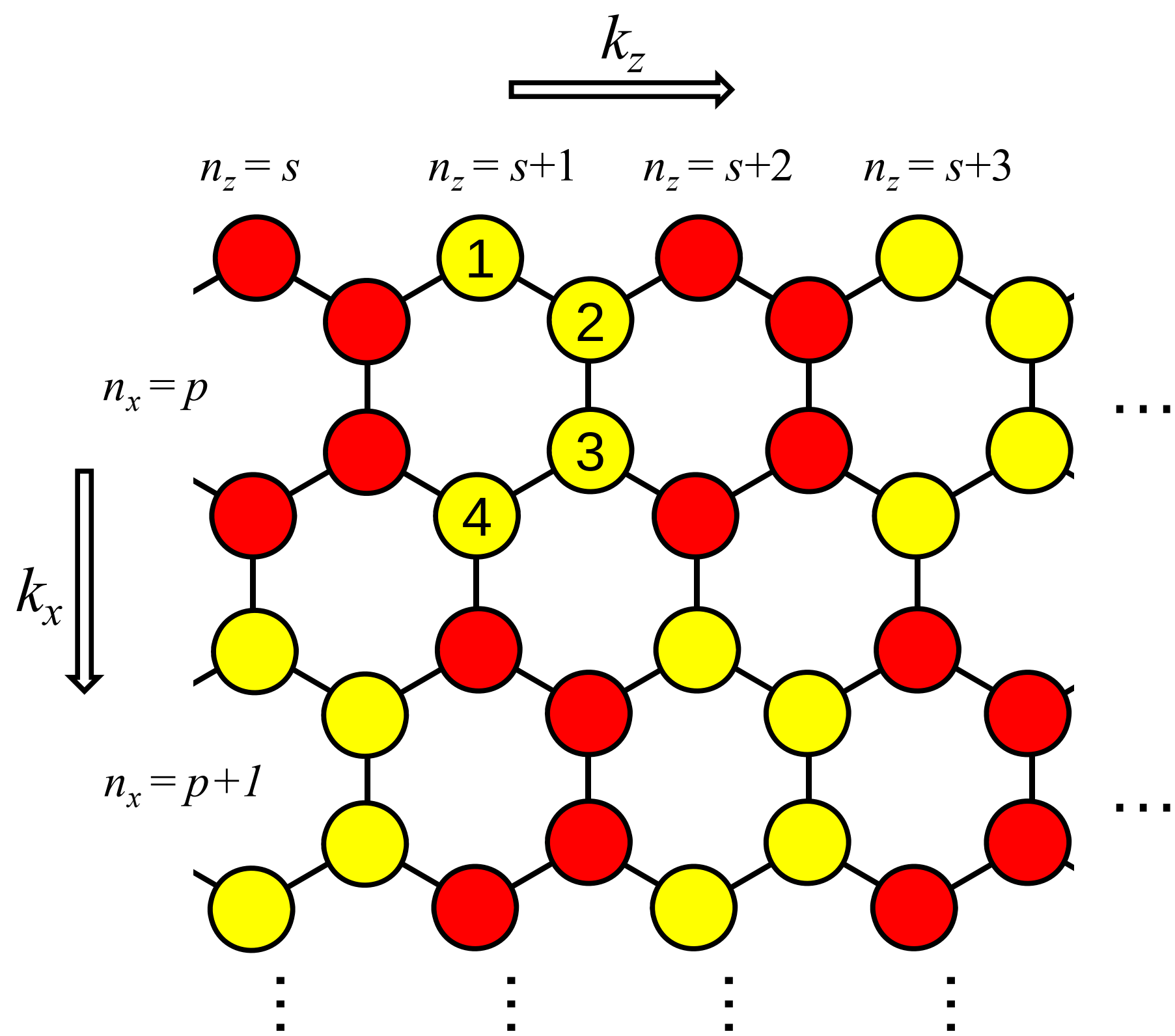

FIG 5. (Color online) Schematic of our system of indexing the atoms in the 1920-atom graphene sheet in our computation of the spectral energy density. The lattice is divided into 4-atom unit cells with translational symmetries in the x and z-direction. The number of repeating units in the x and z-direction are 12 and 40 respectively. The translational symmetries also imply that we can define the wave vectors $k_x$ (= 1 to 12) and $k_z$ (= 1 to 40). The reflection symmetry about the x-y plane implies that the $k_z = n$ and the $k_z = 40\text{-}n$ modes are degenerate.